# Two-Dimensional Materials toward CMOS-Compatible Scalable Quantum Hardware

*Mara Liebregts, Ye Wang**

Department of Applied Physics and Science Education, Eindhoven University of Technology, De Groene Loper 19, 5612 AP, Eindhoven, The Netherlands

Corresponding Author: Ye Wang, y.wang19@tue.nl

KEYWORDS

2D materials, CMOS, quantum hardware, scalable, qubit

ABSTRACT

Manufacturability is one of several constraints on developing solid-state quantum processors, along with qubit performance, control, connectivity, cryogenic operation, and fault tolerance. Chemical residues, structural disorder, and non-uniform interfaces introduced during fabrication can compromise qubit coherence. Two-dimensional (2D) materials, with their atomically thin crystals and van der Waals interfaces, offer routes to address some of these challenges. Their concurrent development as next-generation CMOS channel materials motivates an examination of whether relevant processes can be transferred to quantum chips. This Perspective evaluates the opportunities and limitations of 2D materials for CMOS-compatible quantum hardware.

Demonstrated qubits are distinguished from demonstrated components, demonstrated physical phenomena, and proposed concepts. Barriers between laboratory devices and manufacturable technology are identified, and three testable device concepts enabled by 2D materials are presented. The comparisons point to the specific experiments that would show where 2D materials can contribute to scalable quantum hardware.

## MAIN TEXT

Quantum computers promise to solve certain classes of problems beyond the reach of classical computers. Turning laboratory devices into useful processors requires both reliable qubit operation and reproducible fabrication. The development of 2D transition metal dichalcogenides (TMDs) as channels for advanced CMOS raises a related question: can the processes being developed for these materials also support quantum hardware?[1–5] Silicon spin qubits fabricated in 300 mm foundries already illustrate how industrial processing can contribute to coherent quantum operation.[6,7] For 2D materials, the potential connection is through fabrication and integration processes, whose effects on quantum performance still need to be established.

Figure 1 compares selected CMOS and quantum hardware capabilities without assigning them a common timetable. Semiconductor roadmaps and process demonstrations show increasing activity in 2D materials integration, but most 2D quantum platforms remain supported by individual devices or small device sets. Many of these systems have not yet established comparable performance across laboratories or the technological or economic readiness for scale-up.[8,9] This Perspective therefore examines their quantum performance together with the fabrication and integration requirements that would allow promising devices to be reproduced and combined.

Scalability requires quantum performance, control and readout, fabrication yield, thermal management, and interconnect requirements to remain manageable as the number of devices increases. We assess these requirements at three connected levels. Qubit benchmarks include initialization and readout fidelity, single- and two-qubit gate fidelity, coherence, leakage, crosstalk, and controllable coupling. Manufacturing benchmarks include wafer uniformity, device variation, functional yield, and reproducibility across devices and laboratories. At the system level, calibration, connectivity, cryogenic wiring and power, and fault tolerance overhead must be considered together. The required values depend on the architecture and error-correction scheme; no single device or wafer-process metric determines scalability.[6,7,9–11]

CMOS compatibility depends on the chosen integration route. Front-end-of-line (FEOL), back-end-of-line (BEOL), wafer bonding, and monolithic 3D integration impose different constraints on contamination, lithography, wafer handling, and thermal budget. The connection to CMOS lies in process modules that may be adapted, including growth, transfer, deposition, lithography, etching, gate stacks, contacts, and metrology. Each module must be qualified for its effect on charge noise, microwave loss, coherence, device variation, and cryogenic operation.[11–13] Representative industry and research activities supporting **Figure 1** are distinguished by role and demonstrated capability in Supporting **Table S1**.

## 1. Why Two-Dimensional Materials for Quantum Hardware?

Interfaces can limit solid-state qubit performance. Defects in the amorphous oxide barriers of superconducting circuits contribute to decoherence,[14] while charge fluctuations in the dielectrics surrounding semiconductor quantum dots disturb their confinement potentials.[15] These sensitivities motivate the use of 2D crystals with flat basal planes without dangling bonds to engineer the local environment of a quantum device.[16] A clean crystal surface does not, however, guarantee a clean device interface. Material stability and processing determine whether adsorbates, polymer residues, bubbles, wrinkles, strain, or charged defects remain in the assembled structure.[17–20] The appeal of 2D materials therefore rests on the quantum states they support and on the ways their layers and interfaces can be combined and controlled.

Confinement and interlayer interactions provide access to quantum states that differ from those of the bulk material. Examples include Ising spin–orbit protection, associated with out-of-plane spin locking and in-plane critical fields far beyond the Pauli limit;[21] spin–valley states that can be resolved in quantum dots;[22,23] and moiré superlattices, whose band structure, correlations, and superconductivity depend on the relative arrangement of two crystals.[24,25] These effects expand the choice of states and couplings available for device design. Their use in quantum information depends on whether those states can be prepared, coherently controlled, and read out, as discussed in Section 2.

Van der Waals (vdW) assembly lets these materials combine with superconductors, semiconductors, magnets, and dielectrics without requiring closely matched crystal lattices.[26] Different layers can therefore perform different functions within a vertical device. Layer-by-layer wafer assembly and large-area integration extend this approach beyond individual flakes,[26,27] although alignment, variation between layers, and interface quality must be controlled across the assembled area. For quantum devices, the key question is whether this control preserves coherence and yields reproducible device performance. The concurrent development of 2D semiconductors for electronics also makes such stacks relevant to integrating quantum devices with CMOS control circuitry.[2–5] Section 3 examines the transfer and processing requirements behind this possibility.

Within an assembled structure, the interface offers additional ways to control material properties without directly modifying in-plane bonding. Intercalation,[28,29] molecular adsorption,[30] electrostatic gating, and proximity to adjacent layers can modify electronic, photonic, magnetic, and superconducting behavior.[31] This flexibility allows device properties to be adjusted through the choice of neighboring layers, inserted species, or applied fields. The operating temperature, however, depends on the quantum state and device used. Selected hBN defect spins have shown optical initialization, readout, and coherent control at room temperature.[32] In contrast, the bilayer-graphene devices and superconducting qubits containing 2D barriers cited here were measured at dilution-refrigerator temperatures.[33–35] The room-temperature result supports a specific opportunity for optically addressable defect spins.

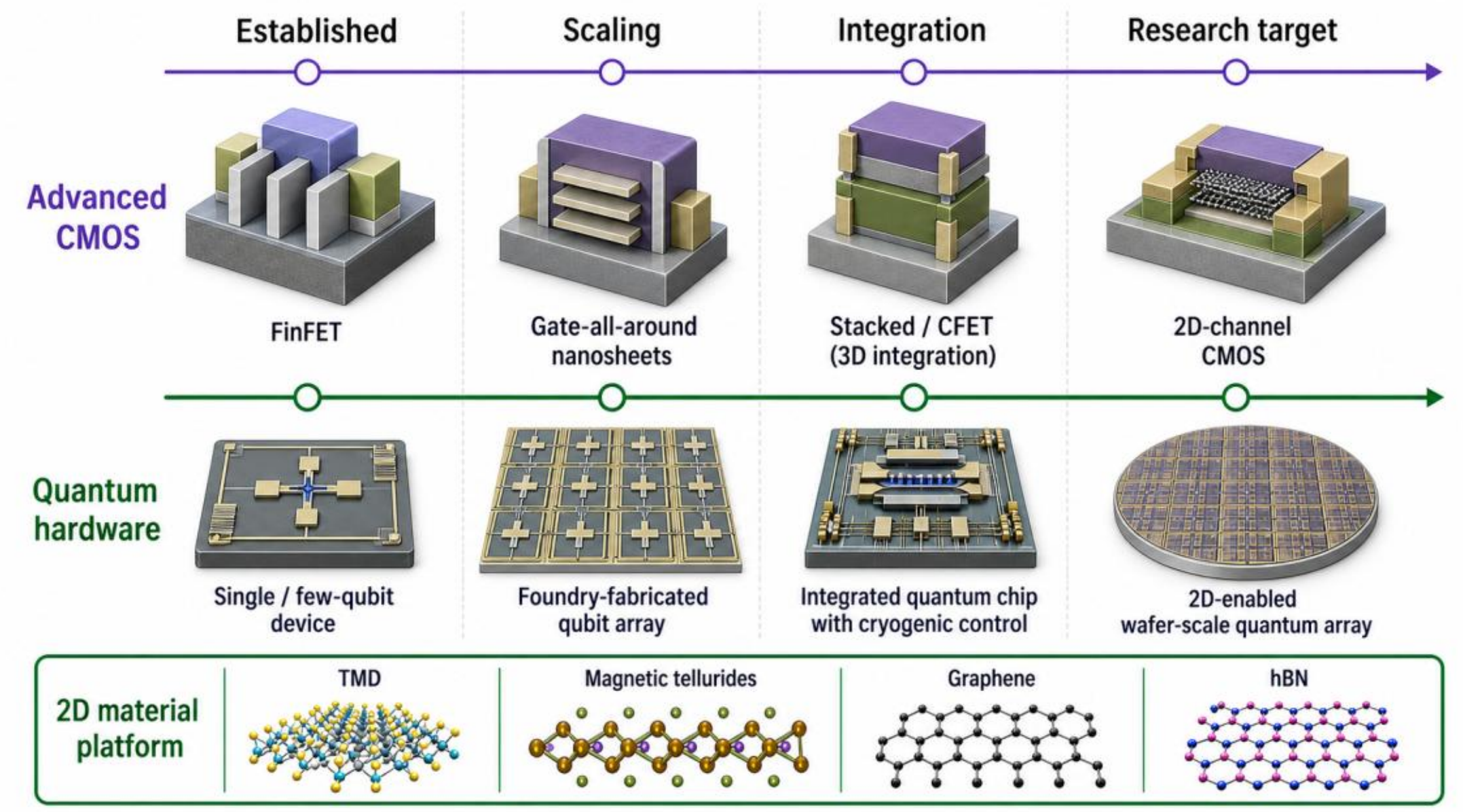

**Figure 1**. Development pathways for advanced CMOS and quantum hardware, and the possible role of transferable 2D material process modules. The upper track follows the progression from FinFETs and gate-all-around nanosheets to stacked and 2D channel devices. The lower track moves from individual qubit devices to larger arrays and integrated control, followed by a research target for 2D-enabled hardware. The stages organize capabilities and research directions; they do not assign equal maturity to the two tracks or predict a common timetable. Foundry arrays and cryogenic control refer to established qubit platforms, not demonstrated wafer-scale 2D qubit arrays. Process modules developed for 2D CMOS may become relevant to quantum hardware if qubit performance, representative yield, and process compatibility are preserved. Supporting Table S1 lists representative activities and their sources.[1,6–9,36]

## 2. Evidence and Readiness of 2D-Enabled Quantum Platforms

Research on 2D quantum materials connects fundamental effects, including Ising superconductivity and moiré correlations, with attempts to build quantum devices. Much depends on reading each experiment at the level it actually establishes. Table 1 and Figure 2 use four classifications: a demonstrated qubit has experimental evidence of state preparation, coherent control, and readout; a demonstrated component performs a relevant function, such as Josephson coupling or single-photon generation; a demonstrated physical phenomenon establishes the underlying effect; and a proposed concept has not yet been implemented in the suggested form. These distinctions allow progress in materials and components to be considered alongside operating qubits without treating them as equivalent demonstrations.

### 2.1 Superconducting Qubits

Superconducting qubits encode quantum information in the two lowest energy levels of an anharmonic microwave oscillator. A Josephson junction provides the nonlinear inductance, while the shunt capacitance sets the charging energy; together, they determine the transition frequency and anharmonicity. In conventional aluminum tunnel junctions, two superconducting electrodes are separated by a thermally grown amorphous $AlO_x$ barrier approximately 1–2 nm thick, through which Cooper pairs tunnel coherently. Disorder in this barrier can host two-level-system defects that contribute to decoherence.[14,37] This sensitivity to the junction and its surrounding dielectric motivates the use of 2D materials, both to modify the interfaces in existing circuits and to introduce junctions with additional means of control.

The first approach replaces conventional junction interfaces or insulating layers with crystalline layers and vdW interfaces. Researchers have formed Josephson junctions via vdW contact between $NbSe_2$ crystals (Figure 2(a)) and incorporated hBN parallel-plate capacitors into transmons (Figure 2(b)).[16,38,39] These implementations address different parts of the circuit. The $NbSe_2$ devices establish Josephson transport across a vdW contact, providing component-level evidence. The hBN capacitor transmons provide circuit-level evidence that a compact crystalline capacitor can support microsecond-scale qubit coherence.

The second approach uses the electronic tunability of 2D materials to control the Josephson coupling. Graphene junctions have enabled gatemon qubits in which an electrostatic gate tunes the junction, providing an alternative to magnetic-flux control.[40] In magic-angle bilayer graphene, this control extends to the formation of the junction itself: gates tune neighboring regions of a continuous crystal into superconducting and normal states, defining both the electrodes and the weak link without a separate barrier layer.[41,42] Twist provides another way to modify the coupling, as illustrated by twisted cuprate junctions exhibiting time-reversal symmetry breaking.[43] Unlike gate voltage, however, the twist angle cannot be adjusted after assembly, making angular accuracy and reproducibility important fabrication requirements. These control methods have reached different stages of development. Graphene gatemons, like hBN capacitor transmons, have been operated as qubits; the moiré and twisted cuprate studies discussed here establish junction behavior or related physical effects.

## 2.2 Semiconductor Spin Qubits

Semiconductor spin qubits store quantum information in the spin state of electrostatically confined charge carriers. In monolayer TMDs and bilayer graphene, the hexagonal lattice provides two inequivalent band extrema, at the K and K′ points, and spin–orbit coupling ties the spin to this valley index, making the valley degree of freedom part of the quantum-dot spectrum. Gate-defined bilayer-graphene quantum dots resolve the spin–valley spectrum down to the last electron.[22,23] Later experiments report spin relaxation up to 60 ms and valley relaxation exceeding 500 ms in a double dot.[44] A single-hole device exhibits a spin–valley relaxation time of 38 s at 30 mK and Kramers-doublet single-shot readout above 99% fidelity.[35] Coherent charge oscillations have also been reported in a bilayer-graphene double dot, with charge dephasing times of approximately 400–500 ps.[33]

The electrostatic confinement used in these devices also connects to CMOS processing: gate structures developed to control a transistor channel can be adapted to confine individual carriers. Silicon spin qubits already exploit this resemblance on industrial 300 mm CMOS platforms.[6,7,45,46] For 2D materials, a similar process overlap is emerging: $MoS_2$, $WS_2$ and $WSe_2$ are under industrial development as logic channels, using related growth methods, gate stacks and dielectrics.[2–5] Spin qubits may therefore benefit particularly from transferable modules in the classical 2D roadmap, although quantum-level performance and yield must be demonstrated independently.

## 2.3 Optically Addressable Defect Spins and Photonic Qubits

Optically addressable spin qubits couple a localized spin to an optical transition, allowing initialization and readout by light and offering an interface to photonic quantum states. Selected hBN colour centres have demonstrated these functions, along with coherent spin control, at room temperature.[32,47–51] For one selected spin, the reported relaxation time is $T_1 = 16.17 \pm 1.55$ μs and Hahn-echo coherence is $T_2 = 2.45 \pm 0.41$ μs.[32] Figure 2(d) shows optical and magnetic resonance characterization of hBN defects; the coherent single-spin control experiment is reported separately in ref 32. Strain-induced emitters in $WSe_2$ provide a complementary photonic component by generating single photons at lithographically defined positions.[52]

The thin host is useful for both sensing and optical integration. A defect in an atomically thin layer can be brought close to a target, while the host can be transferred onto a waveguide or cavity without lattice-matched growth. These possibilities distinguish layered hosts from bulk hosts such as diamond, but proximity to a surface also exposes the center to charge noise and adsorbates. Practical integration must therefore preserve optical and spin properties through transfer, with spectral variability and usable-center yield measured across devices.

Table 1. Maturity, reported quantum evidence, and statistical basis of selected 2D platforms, with 300 mm silicon and superconducting reference devices. The classification applies to the evidence cited in each row. NR means that the specified metric is not reported in those studies; it does not mean zero performance or prove absence from all literature. Relaxation, coherence, gate fidelity, electrostatic yield, and functional-qubit yield measure different properties and are not interchangeable. Reference devices are included as benchmarks, not as 2D material platforms. Proposed concepts have no measured qubit yield.

| Platform | Maturity | Quantum evidence | Scale and statistics | Next requirement |
|---|---|---|---|---|
| hBN capacitor transmon | demonstrated qubit | Coherence up to 25 μs; low-temperature dielectric benchmark.[16,38] | Individual devices; wafer coherence and yield distributions NR. | Large-area hBN with reproducible loss, gate fidelity, and yield. |
| Graphene vdW gatemon | demonstrated qubit | Gate-tunable junction with coherent control.[40] | Individual devices; wafer fidelity and yield distributions NR. | Reproducible junctions and quantum performance after fabrication. |
| $NbSe_2$-based vdW Josephson junction | demonstrated component | Supercurrent, hysteresis, and Fraunhofer response.[39] | Individual junctions; wafer quantum-performance distributions NR. | Junction variability, microwave loss, and coherent qubit operation. |
| Gate-defined moiré junction / SQUID | demonstrated component | Josephson transport and electrostatic control.[41,42] | Laboratory devices; wafer twist and quantum-yield statistics NR. | Control of twist and strain, followed by qubit operation. |
| Twisted-cuprate junction | demonstrated component | Josephson behavior consistent with time-reversal symmetry breaking.[43] | Assembled junctions; wafer qubit statistics NR. | Reproducible circuit behavior and coherent control. |
| Bilayer-graphene quantum dot | demonstrated component | Spin–valley spectroscopy; long relaxation and single-shot readout.[22,23,35,44] | Individual devices / small sets; wafer coherent-spin yield NR. | Coherent spin control, entangling gates, and gate-fidelity distributions. |
| Selected hBN defect spin | demonstrated qubit | Room-temperature coherent control; $T_1$ = 16.17 ± 1.55 μs; echo $T_2$ = 2.45 ± 0.41 μs.[32] | One selected spin for these values; wafer usable-spin yield NR. | Deterministic identity and placement, optical stability, and reproducibility. |
| Site-controlled $WSe_2$ emitter | demonstrated component | Single-photon emission at patterned positions.[52] | Emitter arrays; quantum-state-control statistics NR. | Spectral uniformity, indistinguishability, and integration yield. |
| Ising pairing/moiré correlated states | Demonstrated physical phenomenon | Superconducting or correlated-state evidence.[21,24,25] | Material studies; no qubit demonstration in these cited reports. | Qubit encoding, initialization, control, and readout. |

| Platform | Maturity | Quantum evidence | Scale and statistics | Next requirement |
|---|---|---|---|---|
| Intercalated-ion qubit | proposed concept | Rare-earth control in other hosts and vdW insertion chemistry.[28,53,54] | Proposed architecture; qubit data not applicable. | Site and charge-state control, coherent operation, and coupling. |
| Chiral-junction qubit | proposed concept | Chiral transport and superconducting ingredients.[29,31,55–59] | Proposed architecture; qubit data not applicable. | Zero-field $\varphi_0$ response followed by coherent circuit operation. |
| Moiré multilevel qudit | proposed concept | Twist-dependent bands and correlated states.[24,25] | Proposed architecture; qudit data not applicable. | Addressable levels, coherence, readout, and twist reproducibility. |
| 300 mm Si quantum-dot array reference | demonstrated component | Wafer electrostatics; these yield values are not coherent-qubit fidelity.[10] | 232 twelve-dot devices; 99.8% dot yield; 96% full-device yield; random threshold variation 59 mV. | Connect fabrication yield to coherent-qubit distributions. |
| 300 mm Si two-qubit reference | demonstrated qubit | Measured one- and two-qubit gate fidelities >99%.[6] | Four selected two-qubit devices; not a wafer-wide fidelity distribution. | Representative array-wide gate performance and yield. |
| 300 mm Si eight-qubit reference | demonstrated qubit | All eight addressed; entangling operation on one adjacent pair.[7] | One eight-qubit array; full-array entangling-gate yield NR. | Reproducible entangling operation across the array. |
| 300 mm superconducting reference | demonstrated qubit | Functional transmons measured cryogenically.[9] | 393/400 functional qubits (98.25%); 12,840 junction test structures; inferred frequency RSD 5–7%. | Preserve operation in coupled processors; device yield is not processor yield. |

## 2.4 Cryogenic Control with 2D Electronics

A quantum chip also needs classical electronics for control and readout, so researchers are exploring 2D transistors.[60] Communicating with room-temperature instruments through extensive cabling introduces heat loads and wiring constraints that become harder to manage as qubit number grows.[61] Moving some control circuitry to cryogenic stages can shorten this path,[36] but the circuit must fit the noise and cooling budget of the stage where it operates. At the mixing chamber, available cooling power is limited to tens to hundreds of microwatts, depending on the operating temperature and the refrigerator.[61]

2D field-effect transistors are candidates for such circuitry because electrostatic carrier control can avoid some of the freeze-out associated with chemical doping. At the same time, suitable wide-bandgap channels can support low off-state leakage.[60] Their gain, noise, speed, variability, power, and yield still need to be compared with silicon cryo-CMOS at the intended temperature. vdW stacking could place these transistors above or below a qubit layer, shortening the control path, provided that electrical isolation, alignment, and thermal crosstalk remain acceptable.

Control becomes more demanding when individually operable qubits are coupled into an array. In a recent 300 mm foundry-fabricated SiMOS device, all eight qubits were tuned and addressed, with $T_2^*$ up to 41(2) μs and Hahn-echo $T_2$ up to 1.31(4) ms. An entangling operation was demonstrated for one adjacent pair, leaving reproducible entangling-gate tuning across the array as a further requirement.[7] The example shows why fabrication capability must be considered together with coupling, crosstalk, leakage, and automated calibration, as well as the wiring, power, and fault tolerance demands of the full processor.[11,36,61] Table 1 separates this coherent-device evidence from wafer-level fabrication statistics.

Read together, the platforms above point to four measurements that would each move a platform to its next evidence tier. First, for the vdW and twisted junctions, which so far establish junction behavior, microwave spectroscopy and coherent control of the junction itself; qubit operation has been reported for the hBN-capacitor transmon and the graphene gatemon. Second, for the bilayer-graphene spin and spin–valley dots, which have measured relaxation and single-shot readout, spin coherence, coherent spin control, and entangling-gate fidelity. Third, for the $WSe_2$ emitters, which have established single-center optical properties, encoding, manipulation, and readout of a photonic qubit; for the selected hBN defect spins already operated as qubits, reproducible preparation and consistent optical and spin properties across centers. Fourth, for the cryogenic 2D electronics, a system-level comparison with silicon cryo-CMOS at the intended temperature.

Across all four, the reported values come from one or a few devices, so distributions and yield are common requirements. Sections 3 and 4 treat this list as the measurement agenda.

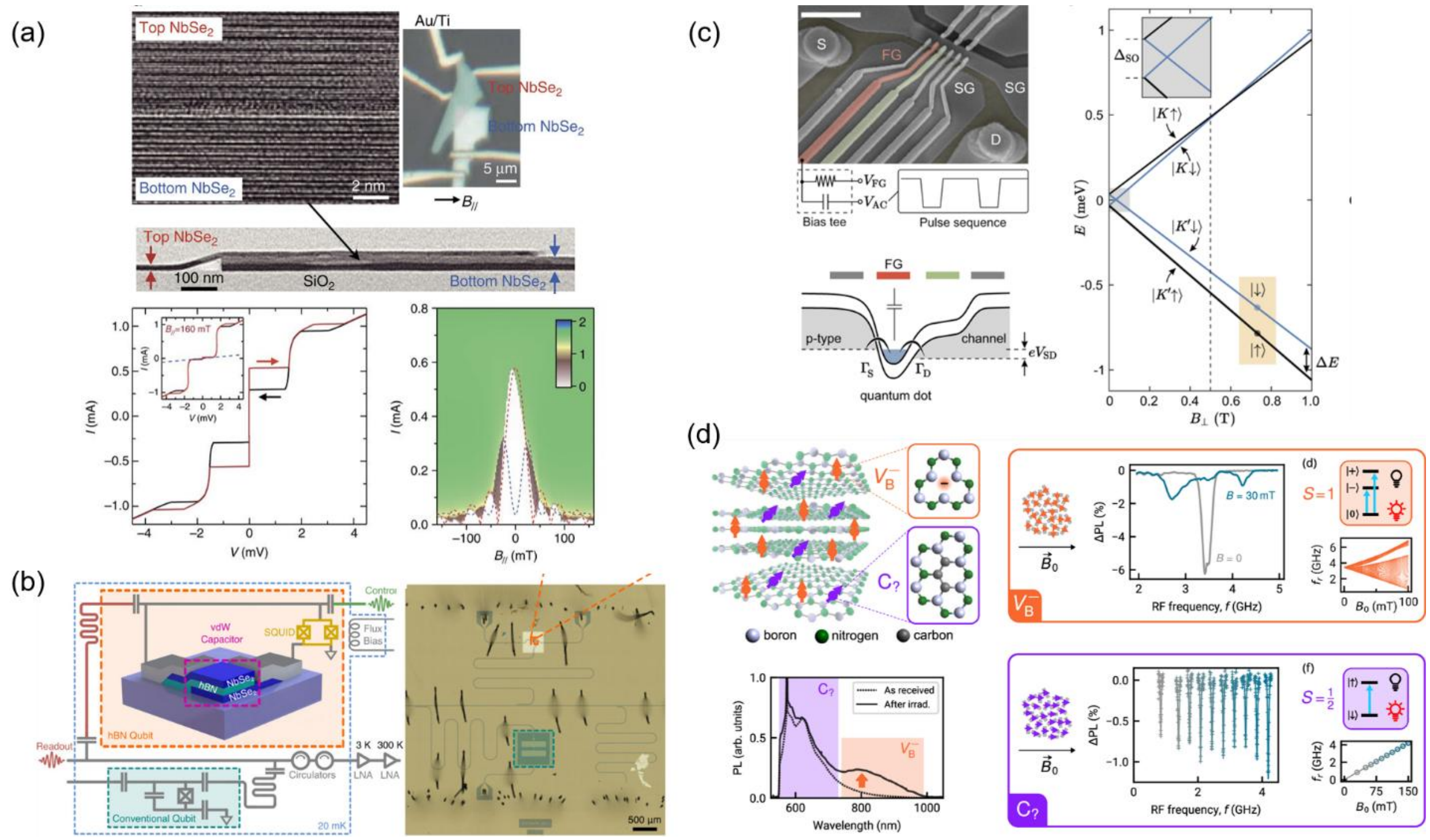

**Figure 2**. Representative 2D qubits, components, and physical phenomena, classified by the evidence shown. (a) A vdW Josephson junction formed by vdW contact between two $NbSe_2$ crystals, showing hysteretic current–voltage response and a Fraunhofer critical-current pattern (demonstrated component). (b) A miniaturized transmon whose shunt capacitor is a $NbSe_2$/hBN/$NbSe_2$ stack, coupled to SQUID, control and readout lines at 20 mK (demonstrated qubit). (c) A gate-defined single-electron quantum dot in bilayer graphene, with the perpendicular-field spectrum resolving K/K′ valley states split by spin–orbit coupling and Zeeman energy (demonstrated component). (d) Room-temperature spin defects in hBN: the S = 1 boron vacancy and an S = 1/2 carbon-related defect, characterized by photoluminescence and optically detected magnetic resonance (demonstrated physical phenomenon). The spectra in panel (d) should be distinguished from coherent control of selected hBN spins reported elsewhere.[32] Panels (a), (c) and (d) adapted from refs [39], 23 and [50], respectively, under a Creative Commons CC BY 4.0 license. Panel (b) adapted with permission from ref [38]. Copyright 2021 American Chemical Society.

## 3. Manufacturing Barriers

The devices in Section 2 show what the best available crystals and carefully assembled interfaces can achieve. Many still rely on manual exfoliation and flake-by-flake stacking, making it difficult to produce enough comparable devices to separate material variation from device physics. Moving toward manufacturing requires reproducible material supply, patterning and integration suited to large substrates, and metrology that connects process variation to quantum performance. The following sections examine these requirements, including the additional constraints of transfer to a 300 mm process line.

### 3.1 High-Quality Materials Supply: From Bulk Crystals to Wafers

Relatively few laboratories supply high-quality starting crystals. Many devices use hBN exfoliated from crystals grown under high pressure and temperature,[62] together with superconducting or semiconducting layers obtained from a small number of bulk-crystal sources. When performance is measured on only one or a few assembled devices, as in many entries in Table 1, flake-to-flake variation is difficult to distinguish from differences in device design or processing. Comparisons between laboratories become meaningful only with reproducible crystal supply and measurements across material batches.

Chemical vapor deposition (CVD) and metal-organic chemical vapor deposition (MOCVD) have advanced large-area growth,[63–67] with much of this development directed toward semiconducting and dielectric films for classical transistors. Progress also extends to the material classes considered here (Figure 3(a)). 2D superconductors have advanced from micrometer-scale $Mo_2C$ and $NbSe_2$ crystals to wafer-scale single-crystal films,[68–71] and 2D ferromagnets have reached wafer scale through molecular-beam epitaxy and interfacial engineering.[72–75] hBN growth now includes large-roll monolayers, wafer-scale single crystals, and controlled multilayers,[51,76–80] while centimeter-scale carbon-doped films have demonstrated room-temperature quantum emission.[48] These increases in synthesis scale create opportunities for more representative device studies; a supply chain with quantum performance established across material batches is the step after that.

Achieving this consistency requires control of nucleation, stoichiometry, and thermal history. Misoriented islands can leave grain boundaries when they coalesce, and insufficient chalcogen supply during growth or cooling can introduce vacancies. Air-sensitive crystals also need protection during handling. How strongly these imperfections affect a quantum device depends on its function, so evaluate growth quality against the requirements of the intended platform.

The integration route imposes separate constraints on growth and processing temperature. FEOL integration may allow higher-temperature steps before sensitive device and interconnect layers form. BEOL and monolithic 3D integration generally allow less thermal exposure because completed transistors, contacts, dielectrics, and interconnects must be preserved. Wafer bonding can separate high-temperature growth from final integration, but it adds cleanliness, alignment, and bonding-yield requirements. Approximately 400 °C is a commonly used BEOL guideline. The insertion point, peak temperature, dwell time, ambient, and existing material stack must be specified for each process.[11–13]

Material specifications must also reflect how a given device stores and controls quantum information. Uncontrolled defects can introduce charge noise or dielectric loss, whereas a deliberately introduced defect may itself provide the spin used as a qubit. Any definition of "quantum-grade" material must therefore be platform-specific. $MoS_2$ defect spectroscopy, for example, links vacancy-related states to charge switching and noise, but does not establish a general qubit requirement of $10^{10}$ $cm^{-2}$.[81] Material quality must be assessed through its effect on the intended device under the relevant operating conditions.

For spin and spin–valley quantum dots, the first requirement is a sufficiently stable electrostatic environment to confine and control individual carriers. Consider charge noise and traps in the host material or gate dielectric alongside the energy levels that define the qubit. Orbital and valley splittings, intervalley mixing, spin–orbit coupling, and spin-carrying isotopes influence which states can be addressed and how they relax or lose coherence. In bilayer graphene, single-electron spectroscopy resolves a spin–orbit gap of about 60 μeV and places an experimental upper bound of 20 μeV on intervalley mixing.[22] These results illustrate how spectroscopy can test whether the intended states remain distinguishable. For TMD dots, the corresponding assessment must examine how vacancy-related traps and atomic-scale disorder contribute to charge noise and intervalley scattering. These quantities connect to spin coherence, readout, and gate fidelity measured across multiple devices.[11,15,35,46]

Superconducting circuits place different demands on the same materials because performance depends on microwave loss and Josephson-junction properties. Relevant measurements include dielectric loss at the operating temperature and photon occupation, loss associated with two-level systems, and variation in junction properties. An hBN capacitor study bounds the low-temperature, single-photon loss tangent to the mid-$10^{-6}$ range and reports transmon coherence reaching 25 μs.[16] This provides a benchmark for that capacitor and circuit geometry. The surfaces and interfaces surrounding the capacitor also contribute to the measured loss.[14,37]

For optically addressable defect spins, the objective is not to reduce the total number of defects: the desired center must be created reproducibly while limiting unwanted defects and nearby noise sources. Material assessment must distinguish the center's identity, charge state, placement, and concentration from parasitic disorder. These properties must then be related to optical stability, spin initialization and readout, relaxation, coherence, and the yield of usable centers. The room-temperature results discussed in Section 2.3 demonstrate the properties of an individual spin; population measurements would be the next step toward reproducible material specifications.[32,47,50]

What the three platforms have in common is the form a useful specification must take. It is a distribution measured across devices, tied to the quantity that platform actually depends on: spin coherence and gate fidelity for the quantum dots, loss tangent and junction spread for the superconducting circuits, and usable-center yield for the defect spins. Single-number material metrics, whether a vacancy density or a spectroscopic gap from one device, do not substitute for these.

Because the requirements differ in this way, the synthesis target differs with them. For devices based on an established host material, the priority may be to reduce unwanted disorder or introduce a controlled population of useful defects. Other proposed devices require access to a particular phase or stacking configuration, so improving the purity of the starting crystal alone is insufficient. **Figure 3(b)** illustrates this second challenge: a thermodynamically stable phase may be readily accessible, while a desired metastable phase remains difficult to reach or reproduce because of kinetic barriers. Adjusting temperature, pressure, and precursor ratios may not provide sufficient control over the pathway to that state.

Growth and post-growth processing offer complementary ways to address this problem. Mechanochemistry, interfacial epitaxy, and twist epitaxy provide routes for modifying reaction pathways, phase formation, or stacking during synthesis and assembly.[82–84] After growth, surface functionalization,[30] intercalation,[28,29] and phase engineering[56] offer ways to modify crystals without requiring the desired functionality to emerge during the initial growth step. When these treatments preserve crystal and interface quality, they could separate host-material production from the introduction of quantum functionality. Device-specific measurements described above, together with reproducibility across material batches and fabricated devices, will determine whether they help scalable quantum hardware.

### 3.2 Industrially Compatible Patterning and Integration

High-quality materials are only the starting point; working quantum devices depend on nanofabrication. Lithographic patterning sets interface quality and determines manufacturing reproducibility and reliability. At present, 2D quantum devices are patterned predominantly by electron-beam lithography (EBL), a serial direct-write technique whose low throughput limits its use in high-volume wafer manufacturing. High-throughput fabrication on 300 mm substrates favors projection methods, including deep ultraviolet and extreme ultraviolet (EUV) lithography. An institutional announcement reports High-NA EUV patterning of a quantum-dot device with 6 nm gate gaps in a 300 mm fab-compatible process.[45] An EUV scanner operates in a low-pressure hydrogen background that 13.5 nm radiation can excite into a transient plasma.[85,86] Studies on semiconducting TMDs indicate that EUV exposure and the accompanying photo-electrons and hydrogen-plasma species break Mo–S bonds in $MoS_2$, generating sulfur vacancies and hydrogen functionalization that raise the defect density and modify carrier transport (Figure 3(c)).[87,88] Superconducting TMDs such as $NbSe_2$ could undergo analogous chalcogen-vacancy chemistry, with consequences for the superconducting order that are unknown.

Contact formation requires a related change in processing. Laboratory devices commonly use transferred graphene contacts or evaporated metals followed by solvent-based lift-off. Foundry production instead deposits metal films and patterns them by etching through a lithographic mask, which requires low-damage, high-resolution, uniform deposition and removal. Atomic-layer etching of $WSe_2$ and $MoS_2$ demonstrates controlled layer-by-layer removal,[89,90] but quantum materials will require dedicated chemistries. Classical 2D CMOS integration faces the same damage and contact problems, so low-damage etch and deposition chemistries developed there are candidates for adaptation.

Transfer and stacking introduce further sources of variation after growth. Residues, tears, wrinkles, trapped contamination, bubbles, particles, strain, alignment errors, and local delamination can remain at the interfaces created by polymer-assisted or mechanical transfer.[17–20] Their distribution matters as much as the appearance of one clean region. Wafer studies should report areal defect density, alignment distributions, bonded area fraction, variation between layers, and device yield over the full assembly sequence, including losses from repeated stacking. Correlating these measurements with microwave loss, charge noise, coherence, or spectral stability would show which process variations limit quantum performance. Large-area transistor demonstrations provide useful starting points for this assessment.[26,27]

Recent transfer methods have addressed some of these limitations. Polymer-free silicon nitride membranes allow assembly without an organic support contacting the active layers, improving interface cleanliness and moiré uniformity.[18] Wafer-scale vdW integration has also demonstrated reproducible transistor fabrication with reduced bubble and wrinkle formation.[27] Water-based transfer has been demonstrated from 100 mm sapphire substrates, although the method still uses a PMMA support.[91] Mica-assisted assembly provides another polymer-free route with clean interfaces and preserved twist angle; its current implementation is best suited to small-batch research devices up to approximately 200 μm in lateral size.[92] These methods address different scales and sources of contamination. Comparing their resulting device distributions, using the measurements outlined above, would help determine which routes are suitable for quantum integration. In each case the qualifying data are quantum-device data: an etch chemistry, a transfer route, or a lithography step carried over from classical 2D CMOS must be requalified against coherence, loss, and yield measured on the quantum device itself.

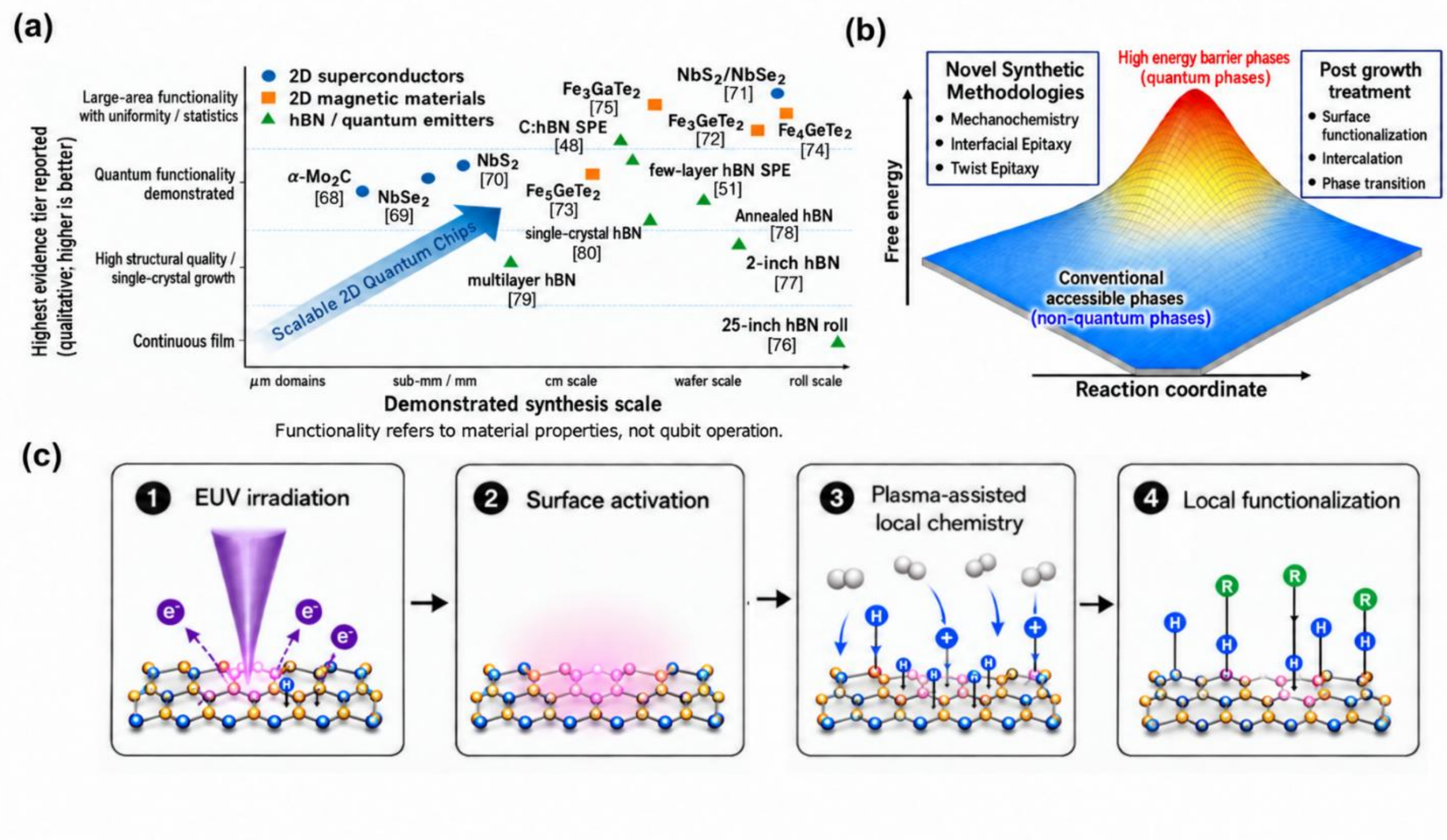

**Figure 3**. Manufacturing barriers to scalable 2D quantum chips. (a) Highest reported evidence tier (continuous film → single-crystal growth → demonstrated quantum functionality → large-area functionality with statistics) versus demonstrated synthesis scale for 2D superconductors (refs 68–71), 2D magnets (refs 72–75) and hBN or quantum emitters (refs 48, 51, 76–80). The diagonal arrow marks the route toward scalable 2D quantum chips. Functionality refers to material properties, not qubit operation; the evidence tiers do not establish wafer-scale qubit yield or coherent control. (b) Free-energy landscape of 2D quantum-material synthesis: conventional phases lie in an accessible basin, whereas target phases beyond a high barrier require directional routes such as vdW mechanochemistry, interfacial epitaxy or twist epitaxy. (c) Possible interaction of EUV exposure with a 2D material; the chemistry and consequences for quantum properties remain to be established.

## 3.3 Metrology and Smart Manufacturing

The preceding examples show why metrology must connect fabrication records to device performance. CMOS process development relies on stable process windows and inline acceptance criteria; for quantum hardware, those criteria must capture variations that affect charge noise, microwave loss, coherence, frequency, and yield.[9–11,37,46] Many of these effects become apparent only at cryogenic temperatures, too late for room-temperature inspection alone to guide acceptance. Linking wafer maps to cryogenic measurements can identify which material and process variations matter, allowing subsequent runs to be adjusted on the basis of device populations instead of operator-specific trial and error.

Figure 4 organizes this connection into a three-stage metrology loop. Stage one uses room-temperature tools, including Raman and photoluminescence mapping, scanning-probe measurements of interfaces, and four-point-probe measurements, to screen wafers before packaging. Stage two adds cryogenic statistical characterization through multiplexed measurement of device populations,[10,93] together with resonator-based two-level-system spectroscopy as an interface diagnostic.[14] Stage three uses the measured distributions and their associated fabrication records to guide the next growth or patterning iteration. Post-fabrication methods such as laser annealing illustrate how measured device variation guides adjustments.[94]

Statistical analysis and machine learning can use these linked datasets to identify correlations between process settings and device properties, detect drift, prioritize experiments, and refine process windows. Models need traceable records, quantified measurement uncertainty, and validation on independent wafers; a correlation on its own cannot identify why a device failed. A recent preprint describes AI-guided recipe development coupled to a semi-automated CVD system with human oversight.[95] It provides an example of feedback in materials research, while qualification for quantum hardware still requires cryogenic measurements and reproducibility across device populations.

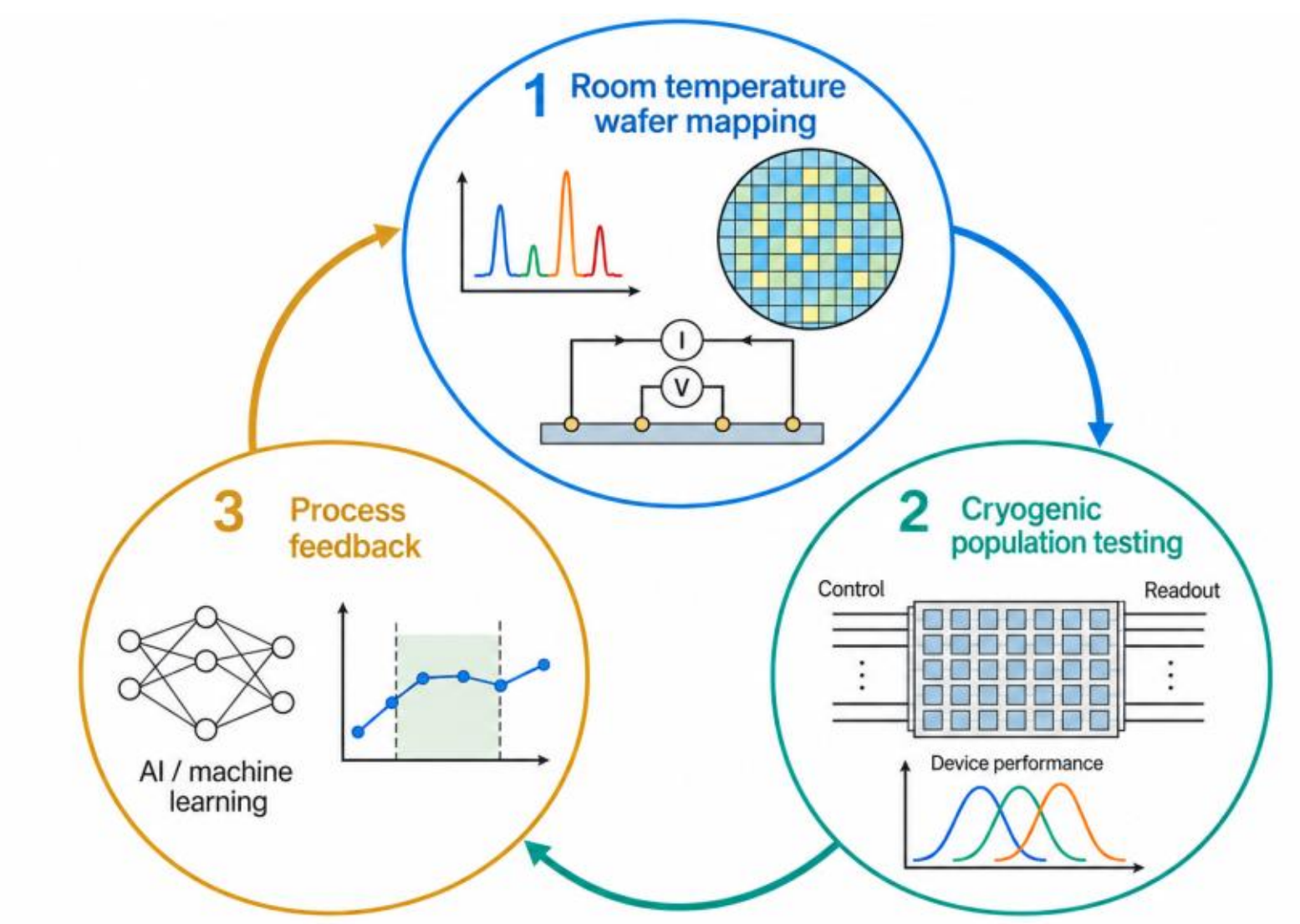

**Figure 4**. Closed qualification loop for CMOS-compatible 2D quantum hardware. Room-temperature wafer mapping (1) screens material and process variation. Cryogenic population testing (2) measures distributions of device and quantum properties. Process feedback (3) links these measurements to growth, transfer, patterning, and treatment records. Statistical analysis and AI or machine learning act on the linked records to flag process drift. Proposed changes must be tested on independent wafers, with quantum performance and yield measured again before process qualification. All maps, distributions, and device drawings are schematic.

## 4. Critical Outlook and Testable Research Concepts

The experimental results surveyed here suggest different next steps for different platforms. Some still require coherent control or reliable readout; others can already operate as qubits but need improved coupling, fidelity, or reproducibility. Producing comparable devices is part of answering these physical questions: without sufficient material and process consistency, it is difficult to determine whether a change in design has improved performance.

The eventual value of a materials improvement must also be assessed at the processor level. Even systems with around 100 qubits require carefully engineered cryogenic wiring and heat management.[61] If crystalline vdW interfaces reproducibly reduce microwave loss and physical error rates, they could reduce the number of physical qubits needed for a given error-corrected task. Whether that also reduces control lines or power depends on the architecture.[36,61] Similarly, the cryogenic 2D electronics discussed in Section 2.4 may enable closer integration through shared materials and selected fabrication steps,[60] but shorter connections would be useful only if their benefits survive the added noise and heat load. These system comparisons are needed to identify where a 2D component offers a practical improvement over an established alternative.

vdW structures also invite qubit designs quite different from those discussed so far. Their interlayer spaces accommodate functional species, and stacking and twist modify electronic states and coupling. **Figure 5** outlines three concepts based on these features. For each, the relevant materials or physical effects have some experimental support, but the proposed quantum device remains to be demonstrated.

**Intercalated-ion qubits.** Rare-earth ions placed in vdW gaps could provide localized, atom-like states within a layered host (**Figure 5(a)**). Their shielded 4f states and established optical and microwave control methods in other materials motivate this possibility.[53,54] Intercalation offers a potential route to introducing such centers: molecular cations, including chiral species, have been inserted into vdW crystals and selectively introduced into individual components of a stack under mild galvanic conditions.[28] Controlled incorporation of an optically or magnetically active rare-earth center has not yet been shown, but these chemical results suggest a route worth testing. The host would first have to provide reproducible ion positions, charge states, and local environments. Optical or microwave spectroscopy and coherence measurements would then determine whether the inserted ions retain useful quantum properties. Controllable ion–ion coupling and reproducible insertion over larger areas would be needed before this approach could be assessed as an array architecture.

**Chiral-junction qubits.** A chiral interlayer offers a possible means of modifying Josephson coupling through the structure of the barrier (**Figure 5(b)**). Chiral layers can break inversion symmetry and exhibit spin-selective transport through the CISS effect.[55] Whether these properties, together with spin–orbit coupling and the required symmetry breaking, can produce a reproducible $\varphi_0$ junction is the central question for this concept. Several results motivate its investigation. Our recent work demonstrated molecular-encoded symmetry breaking in a chiral 2D material and its coupling to the quantum state of light.[56] Charge-to-spin conversion has also been controlled through the chiral charge-density wave of 1T-$TaS_2$ in a vdW heterostructure,[57] and chirality-dependent conversion has been measured in tellurium.[58] More directly relevant to an interlayer barrier, chiral molecules inserted into layered $TaS_2$ form ordered superlattices with spin selectivity,[29] while related hybrid superlattices exhibit an in-plane critical field beyond the Pauli limit and a field-free superconducting diode effect.[31,59] These findings motivate phase-sensitive junction measurements; they do not themselves demonstrate a $\varphi_0$ phase offset. Establishing such an offset reproducibly, while excluding trapped-flux and domain effects, would be the first step. Microwave spectroscopy and coherent control would then be needed to determine whether the junction can function as a qubit.

**Moiré multilevel qudits.** Twist offers a different route to designing the states used for quantum information. Moiré superlattices support narrow minibands and correlated states whose energies depend on twist and the resulting electronic structure (**Figure 5(c)**).[24,25] The proposed qudit would use more than two distinguishable states within a suitably confined moiré system. These states would need to form a stable, individually addressable set of transitions; the presence of minibands alone falls well short of such a system. State preparation, readout, coherence, and coupling would therefore need to be demonstrated before assessing a computational advantage. Twist and strain variation would also have to remain small enough to permit reproducible control across devices. Multilevel encoding could reduce circuit depth for some tasks, but any benefit would need to be weighed against the added demands of control, leakage suppression, and error correction.

For each concept, reproducible demonstration of the proposed quantum function is the immediate goal. Considering scalability at this stage can still inform the choice of host, interface, and control scheme. Those choices should leave a plausible route to combining devices if the initial experiments succeed, without making large arrays the objective before useful and repeatable quantum operation has been established.

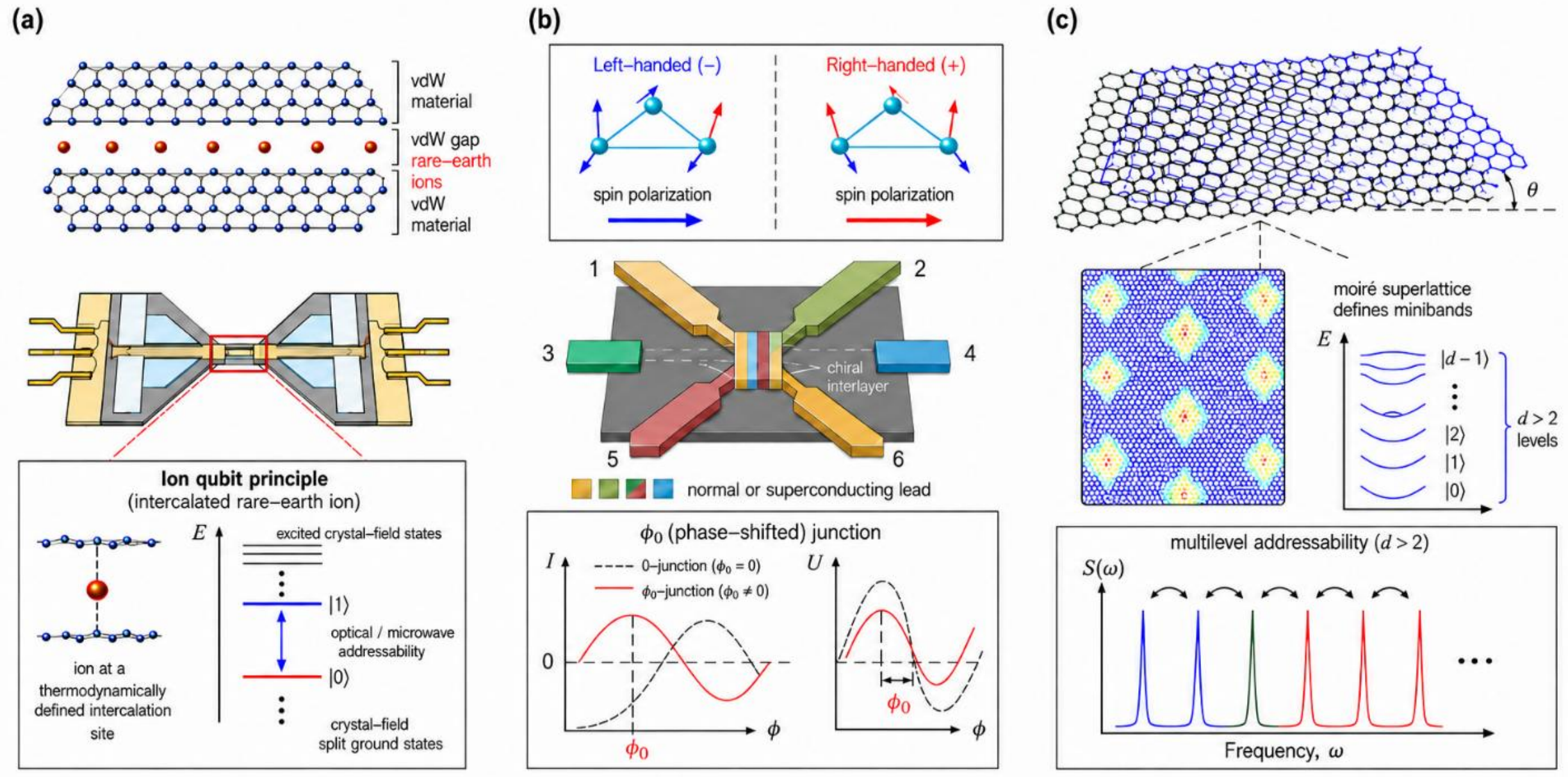

**Figure 5**. Proposed quantum-device concepts enabled by vdW structural degrees of freedom. Each panel is a proposed concept, not an experimentally demonstrated qubit. (a) Intercalated-ion qubits: rare-earth ions ordered within a vdW gap form candidate atom-like sites built into a chip. Controlled site occupancy and charge state, initialization, coherent manipulation, readout, coherence, and ion–ion coupling must be established. (b) Chiral-junction qubits: a CISS-active chiral interlayer could produce a $\varphi_0$ junction whose phase offset is set by structure rather than applied flux. A reproducible zero-field phase shift, with trapped flux and domain effects excluded, should precede tests of anharmonicity, coherent control, and readout. (c) Moiré multilevel qudits: the potential of a twisted bilayer could define a multilevel computational space. Addressable transitions, state preparation and readout, coherence, coupling, and reproducible twist and strain are required. All three concepts also require measurements of device variation and reproducibility.

2D materials add specific functions to quantum hardware without replacing established qubit platforms. Low-loss dielectrics, tunable junctions, defect-spin or photonic interfaces, and potentially cryogenic electronics offer specific functions whose performance can be compared with conventional alternatives. Combining several of these functions in a shared 2D platform remains a longer-term opportunity. The path toward it is through reproducible devices, processing that preserves their quantum properties, and evidence that integration improves performance after control and cooling costs are included. Such comparisons would show which of the interfaces, stacking configurations, and proximity effects available in 2D materials repay their added process cost.

## ASSOCIATED CONTENT

Supporting Information. Representative industry, foundry, research-institute, and consortium activities underlying **Figure 1**, with sources and distinctions between manufacturing, research demonstrations, and roadmap targets (**Table S1**).